\documentclass{aa}
\usepackage{overpic}
\usepackage[fleqn]{amsmath}
\usepackage{graphicx}% Include figure files
\usepackage{dcolumn}% Align table columns on decimal point
\usepackage{bm}% bold math
\usepackage{rotating}
\usepackage{threeparttable}
\usepackage{amsmath}
\usepackage{verbatim}
\usepackage{longtable}
\usepackage{mdwlist}
\usepackage{lscape}
\usepackage{natbib}
\usepackage{multirow}
\usepackage{upgreek}
\usepackage[colorlinks=true, pdfstartview=FitV, linkcolor=blue,citecolor=blue, urlcolor=blue]{hyperref}
\usepackage{wasysym}
\usepackage{txfonts}
\usepackage{gensymb}
\usepackage{morefloats}
\usepackage{footnote}
\begin{document}

\bibliographystyle{plain}

\title{Identification of new M31 star cluster candidates from PAndAS images using convolutional neural networks}
\titlerunning{M31 cluster candidates from PAndAS}
\authorrunning{Wang et al.}

\author{Shoucheng Wang\inst{1,2,3}  \and
          Bingqiu Chen\inst{2} \and
          Jun Ma\inst{1,3} \and
          Qian Long\inst{4} \and
          Haibo Yuan\inst{5} \and
          Dezi Liu\inst{2} \and
          Zhimin Zhou\inst{1} \and
          Wei Liu\inst{6} \and
          Jiamin Chen\inst{7} \and
          Zizhao He\inst{3,8}
          }

\institute{Key Laboratory of Optical Astronomy, National Astronomical Observatories, Chinese Academy of Sciences, Beijing 100012, China e-mail: majun@nao.cas.cn
\and
South-Western Institute for Astronomy Research, Yunnan University, Kunming, Yunnan 650091, P. R. China e-mail: bchen@ynu.edu.cn
\and
School of Astronomy and Space Sciences, University of Chinese Academy of Sciences, Beijing 100049, China
\and
Yunnan Observatories, Chinese Academy of Sciences, Kunming 650216, P. R. China e-mail: longqian@ynao.ac.cn
\and
Department of Astronomy, Beijing Normal University, Beijing 100875, P. R. China
\and
National Laboratory of Pattern Recognition, Institute of Automation, Chinese Academy of Sciences, Beijing 100190, P. R. China
\and
School of Computer Science and Engineering, Central South University, Changsha 410083, P. R. China
\and
Key Laboratory of Space Astronomy and Technology, National Astronomical Observatories, Chinese Academy of Sciences, Beijing 100012, China
}

\abstract
{Identification of new star cluster candidates in M31 is fundamental for the study of the M31 stellar cluster system. The machine-learning method convolutional neural network (CNN) is an efficient algorithm for searching for new M31 star cluster candidates from tens of millions of images from wide-field photometric surveys.}
{We search for new M31 cluster candidates from the high-quality $g$- and $i$-band images of 21,245,632 sources obtained from the Pan-Andromeda Archaeological Survey (PAndAS) through a CNN.}
{We collected confirmed M31 clusters and noncluster objects from the literature as our training sample. Accurate double-channel CNNs were constructed and trained using the training samples. We applied the CNN classification models to the PAndAS $g$- and $i$-band images of over 21 million sources to search new M31 cluster candidates. The CNN predictions were finally checked by five experienced human inspectors to obtain high-confidence M31 star cluster candidates.
}
{After the inspection, we identified a catalogue of 117 new M31 cluster candidates. Most of the new candidates are young clusters that are located in the M31 disk. Their morphology, colours, and magnitudes are similar to those of the confirmed young disk clusters. We also identified eight globular cluster candidates that are located in the M31 halo and exhibit features similar to those of confirmed halo globular clusters. The projected distances to the M31 centre for three of them are larger than 100\,kpc.
}
{}


\maketitle

\section{Introduction}

Star clusters are excellent tracers for the studies of the formation and evolution of their host galaxies. In past decades, many works have been carried out to search and identify new star clusters in our nearest archetypal spiral galaxy M31. We only mention a few of these works. \citet{2000AJ....119..727B} presented a new catalogue of 435 clusters and cluster candidates in M31 using photometric data observed by the 1.2 m telescope of the Fred L. Whipple Observatory and spectroscopic data observed by the Keck LRIS spectrograph \citep{1995PASP..107..375O} and the Blue Channel spectrograph of the Multiple Mirror Telescope. \citet{2004A&A...416..917G} identified 693 clusters and cluster candidates in M31 in the 2MASS database and presented the Revised Bologna Catalogue (RBC) of confirmed and candidate M31 globular clusters. \citet{2004ASPC..327..118H} discovered nine globular clusters in the outer halo of M31 using images from an INT Wide Field Camera survey of the region. Based on the observation data by the imager/spectrograph DoLoRes at the 3.52\,m TNG telescope and the low resolution spectrograph BFOSC mounted at the 1.52\,m Cassini Telescope of the Loiano Observatory, \citet{2007A&A...471..127G} identified five genuine remote globular clusters according to the radial velocities from the spectra of these sources and their extended nature from ground-based optical images. \citet{2010PASP..122..745H} identified 77 new star clusters from the Hubble Space Telescope (HST) WFPC2 observations of active star formation regions of M31. Based on the visual inspection of the image data from the Sloan Digital Sky Survey (SDSS; \citealt{2002AJ....123..485S}), \citet{2013AJ....145...50D} identified 18 new clusters in M31 outer halo. \citet{2014MNRAS.442.2165H} discovered 59 globular clusters and two candidates in the halo of M31 via visual inspection of images from the Pan-Andromeda Archaeological Survey (PAndAS; \citealt{2009Natur.461...66M}). \citet{2015ApJ...802..127J} presented a catalogue of 2753 clusters in M31 using the images from the Panchromatic Hubble Andromeda Treasury (PHAT) survey. \citet{2015RAA....15.1392C} found 28 objects, 5 of which are bona fide and 23 of which are likely globular clusters based on radial velocities obtained with   Large Sky Area Multi-Object Fiber Spectroscopic Telescope (LAMOST; \citealt{2012RAA....12.1197C}) spectra and visual examination of the SDSS images. \citet{2021A&A...645A.115W} classified 12 bona fide star clusters from the cluster candidates in the literature based on their radial velocities obtained with LAMOST spectra.

Most of these previous works identified new M31 star clusters from the images via visual inspection by naked eye. The modern wide-field galaxy surveys have provided us tens of millions of source images. This makes the traditional visual inspection procedures tedious and very time consuming. Recently, one of the machine-learning methods, the convolutional neural network (CNN; \citealt{1980BC...36...193,1998IEEE...86...2278}), has been adopted by astronomers to identify and classify different types of objects and obtain the physical properties of these objects from their images, as the CNNs are particularly suitable for image-recognition tasks. \citet{2017MNRAS.472.1129P} applied a morphological classification method based on CNN to recognize strong gravitational lenses from the images of the Kilo Degree Survey (KiDS; \citealt{2017A&A...604A.134D}). \citet{2019MNRAS.483.4774L} proposed a supervised algorithm based on deep CNN to classify different spectral type of stars from the SDSS 1D stellar spectra. \citet{2019A&A...621A.103B} developed a CNN-based algorithm to simultaneously derive ages, masses, and sizes of star clusters from multi-band images. \citet{2020AJ....160..264B} detected 3380 clusters in M83 from HST observations with a CNN trained on mock clusters. \citet{2020MNRAS.497..556H} applied CNN to images from the KiDS Data Release 3 and identified 48 high-probability strong gravitational lens candidates. Based on a multi-scale CNN, \citet{2021ApJ...907..100P} identified star clusters in the multi-colour images of nearby galaxies based on data from the Treasury Project LEGUS (Legacy ExtraGalactic Ultraviolet Survey; \citealt{2015AJ....149...51C}).

The PAndAS survey, which has a coverage of $>400$ square degrees centred on the M31 and M33, providing us the most extensive panorama of M31 with large projected galactocentric radii. The primary goal of PAndAS is to provide contiguous mapping of the resolved stellar content of the halo of M31 out to approximately half of its halo virial radius (in projection). The coverage of its imaging reaches a projected distance of $R\sim150\,\rm kpc$ from the centre of M31, which makes it possible to identify globular clusters in the remote halo of M31. Based on the data from PAndAS, \citet{2014MNRAS.442.2165H} identified 59 globular clusters and two candidates in the halo of M31 primarily via visual inspection. \citet{2016ApJ...833..167M} presented a comprehensive analysis of the structural properties and luminosities of the 23 dwarf spheroidal galaxies that fall within the footprint of the PAndAS, which represent the large majority of known satellite dwarf galaxies of M31.

In this work, we search for new star cluster candidates in M31, including old globular clusters and young disk clusters, from multi-band images of tens of millions sources released by the PAndAS survey. We have constructed double-channel CNNs and trained the models with empirical samples selected from the literature. The CNN-predicted M31 cluster candidates are finally confirmed by  naked eye to obtain high-probability candidates.

This paper proceeds as follows. In Section~\ref{sec:data} we introduce the data and the selection of the training samples. We describe our CNN models in Section~\ref{sec:cnn} and our search procedure in Section~\ref{sec:search}. Finally, we present our results in Section~\ref{sec:result} and summarize in Section~\ref{sec:summary}.

\section{Data}\label{sec:data}
\subsection{PAndAS images}

The PAndAS survey started collecting data in 2003 and completed the collection in 2010. The survey was conducted in $g$ and $i$ bands using the Canada-France-Hawaii Telescope (CFHT) located on Maunakea, Hawaii. The observations were carried out with the MegaCam wide-field camera \citep{2003SPIE.4841...72B}, which contains 36 2048$\times$4612 pixel CCDs. The effective field of view of the camera is 0.96$\times$0.94\,deg$^{2}$ , with a pixel scale of 0.187$^{\prime\prime}/$pixel. The values of the typical seeing of the PAndAS survey are 0.67$^{\prime\prime}$ and 0.60$^{\prime\prime}$  for the $g$ and $i$ bands \citep{2014MNRAS.442.2165H}, respectively. With the excellent image quality, we are thus able to identify new star clusters in M31. The images were processed with the Cambridge Astronomical Survey Unit (CASU), including image processing, calibration, and photometric measurements. We adopted the processed stacked $g$- and $i$-band images and the merged catalogue of all detected sources as accessed from the PAndAS VOspace$\footnote{\url{https://www.canfar.net/storage/list/PANDAS/PUBLIC}}$.

\subsection{Training samples}\label{sec:augmentation}

\begin{figure*}
\center{
\includegraphics[scale=0.5]{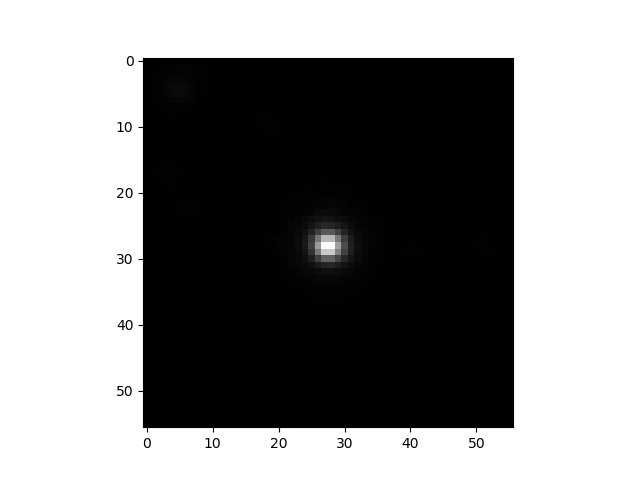}
\includegraphics[scale=0.5]{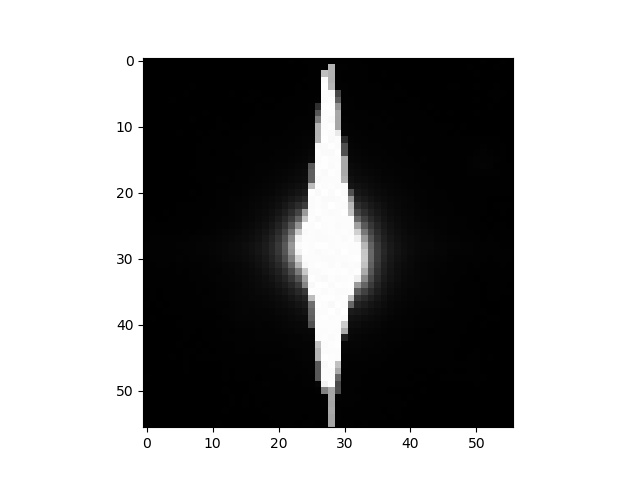}
\includegraphics[scale=0.5]{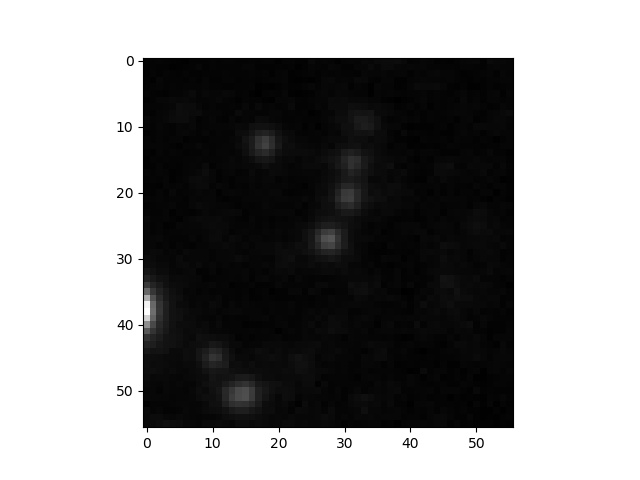}
\includegraphics[scale=0.5]{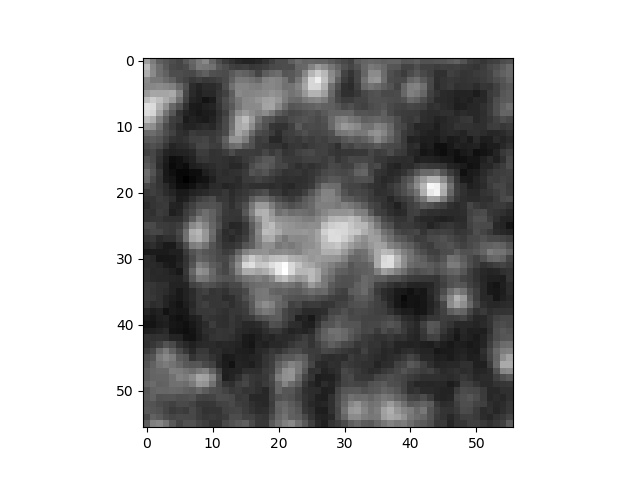}
}
\caption{Example PAndAS images of M31 confirmed star clusters that have been excluded from our positive sample in the current work. From left to right and top to bottom, we show examples of star-like, overexposed, underexposed, and contaminated clusters, respectively. Stamps are centred on the catalogued PAndAS positions and cover 56$\times$56 pixels.}
\label{fig:excluded}
\end{figure*}

To train the CNNs, a large number of positive and negative sample images are required. We adopted empirical training samples. We first collected the confirmed clusters in M31 as the positive training sample. \citet{2021A&A...645A.115W} collected 1233 confirmed clusters from previous works \citep{2004A&A...416..917G,2012ApJ...752...95J,2013AJ....145...50D,2014MNRAS.442.2165H,2015RAA....15.1392C}. We complemented this catalogue with the ``Just star clusters in M31" catalogue contributed by Nelson Caldwell$\footnote{\url{https://www.cfa.harvard.edu/oir/eg/m31clusters/m31clusters.html}}$, which contains 1300 M31 clusters from the literature \citep{2009AJ....137...94C,2011AJ....141...61C,2016ApJ...824...42C}. These two catalogues were merged and cross-matched with the PAndAS source catalogue, which yielded a sample of 1509 confirmed M31 clusters that have been detected by PAndAS. The PAndAS $g$- and $i$-band images of these sources were then collected. We examined these images by naked eye and excluded star-like, overexposed, underexposed, or contaminated images. This selection led to a sample of 975 confirmed M31 star clusters with high-quality PAndAS images. Example images for the sources we excluded are shown in Fig.~\ref{fig:excluded}. We excluded 81, 29, 75, and 349 star-like, overexposed, underexposed and contaminated sources, respectively. Most of the rejected objects are located in a crowded field, that is, the bulge or disk of M31, where there are usually contaminated sources nearby. In addition, most of the brightest ($i<14$\,mag) and faintest sources ($i>23$\,mag) were also excluded. Within the magnitude range of $14<i<23$\,mag, the colour and magnitude distributions of our remaining 975 sample clusters are similar to those of the total catalogue of 1509 clusters, which indicates that there is no selection bias for our positive training sample within the magnitude range of $14<i<23$\,mag.

We cut the image stamps of the remaining 975 confirmed M31 clusters as our positive sample. The size of the samples was 56$\times$56 pixels, which is about 10.47$\times$10.47 arcsec$^{2}$. Because the positive sample was small, we augmented the data to increase the training sample artificially. We flipped the images horizontally or vertically, which yielded 2925 images. The images were then randomly rotated by 90, 180, or 270 degrees, which resulted in 11,700 images. Finally, we randomly shifted the images with $n$ pixels ($-8<n<8$ and $n\neq0$) along the horizontal and vertical directions and obtained 100,400 images of confirmed clusters in M31.

For the negative (non-M31 star cluster objects) samples in our current work, we prepared three catalogues.

(1) Sources that belong to M31, but are not star clusters, including stars, planetary nebulae (PNe), HII regions, and symbiotic stars in M31 were included in the negative sample. For these sources, we adopted the catalogues ``Stars in the M31 catalog'', ``PNe in M31'', ``HII regions in M31'' and ``Symbiotic stars in M31'' contributed by Nelson Caldwell$\footnote{\url{https://www.cfa.harvard.edu/oir/eg/m31clusters/M31\_Hectospec.html}}$ in the current work.

(2) Foreground Galactic sources were included in the negative sample. We collected the Gaia DR2 sources \citep{2018A&A...616A...1G} that are distributed within 2\degr\ to the M31 centre (RA=$00^{\rm h}42^{\rm m}44^{\rm s}.30$, Dec=$+41\degr16^{\prime}09^{\prime\prime}.0$; \citealt{2006AJ....131.1163S}) and have parallaxes higher than 0.01\,mas, that is, distances from the Sun shorter than 100\,kpc.

(3) Background quasars and galaxies were also included in the negative sample. We collected sources from LAMOST DR6 \citep{2015RAA....15.1095L} that have radial velocities higher than $50\,\rm km\,s^{-1}$, which also contain some Milky Way foreground stars.

The catalogues listed above were merged and cross-matched with the PAndAS source catalogue, which yielded a sample of 100,501 non-M31 clusters as our negative sample. Similarly as for the positive sample, we cut the PAndAS images of these sources into 56$\times$56 pixel samples for CNN training.

\section{Convolutional neural networks}\label{sec:cnn}

We used both the PAndAS $g$- and $i$-band images. Thus we constructed double-channel CNNs, which are able to make full use of the morphological information of images in $g$ and $i$ bands.

The CNN is a deep-learning algorithm especially for image and video recognitions. The CNN is able to extract the features of input images and convert them into lower dimensions without loosing their characteristics. A typical CNN contains an input layer and a series of convolutional and pooling layers, which are followed by a full connection layer and an output layer. Table~\ref{tab:architecture} shows our CNN architecture. The first layer is the input layer, which is a double-channel layer fed with image stamps with sizes of $56\times56\times2$ pixels (for both the PAndAS $g$- and $i$-band images). Following the input layer, the second layer is a convolutional layer that aims to maintain the spatial continuity of the input images and extract their local features. The convolutional layer is also a double-channel layer and consists of 64 kernels. Each kernel has a size of 5$\times$5. We used the rectified linear unit (ReLU; \citealt{ReLU2010}) activation function of this convolutional layer, which helps increase the non-linear properties and speeds up the training process considerably. The third layer is a pooling layer that aims to reduce the dimensionality of the data and the parameters to be estimated. It provides translational invariance to the network. It also helps to control overfitting \citep{2010ANN..Springer..P92}. We used max pooling \citep{Maxpooling2010} with a kernel size of 2$\times$2, which takes the maximum value in a connected set of elements of the feature maps. To extract enough features of the training images, we adopted three pairs of convolutional and max pooling layers in the network. The second pair of layers, that is, the fourth and the fifth layers, are similar to the second and third layers, respectively. The third pair (the sixth and seventh layers) are slightly different from the first two pairs. We adopted 128 kernels for the convolutional and the pooling layer. Two fully connection layers (the eighth and ninth layers) followed after the last pooling layer, which causes the nodes to connecting well to each other and accommodates all possible dependences for the entire architecture. Each layer has 1024 neural units. We used the dropout technique to prevent overfitting. Dropout randomly drops units from the neural network during the training process. We set the value of the dropout to 0.5. Finally, the tenth layer is the output layer, which is also a fully connected layer and was set as a softmax function,

\begin{equation}
\begin{aligned}
P_{i}=\frac{e^{i}}{\sum_{j}^{}e^{j}} .
\end{aligned}
\end{equation}

The output value ($e^{i}$) is interpreted into the probability ($P_{i}$) denoting that a source is an M31 star cluster or a non-M31 star cluster.

We used the binary cross-entropy as the loss function of the real and predicted classification to train the network. Similar to \citet{2020MNRAS.497..556H}, we minimised the cross-entropy by adopting an $Adam$ optimiser \citep{2014arXiv1412.6980K} to optimise the network parameters, and especially to minimise the output error. The $Adam$ optimisation algorithm is an extension of the stochastic gradient descent. It introduces a friction term that mitigates the gradient momentum in order to reach faster convergence. We used a batch size of 50 images and trained the network for 80 epochs. The learning rate was set as 0.001.

\begin{table}
\caption{Architecture of our CNN.}
\label{tab:architecture}
\begin{center}
\begin{tabular}{ccccc}
  \hline
  \hline
No. & Layer & Kernel & Features & Activation \\
\quad & \quad & size & (Units) & function \\
\hline
1 & $Input$ & $56\times56\times2$ & 2 & -  \\
2 & $Conv_{1}$ & $5\times5$ & 64 & $ReLU$  \\
3 & $MaxPooling_{1}$ & $2\times2$ & 64 & - \\
4 & $Conv_{2}$ & $5\times5$ & 64 & $ReLU$ \\
5 & $MaxPooling_{2}$ & $2\times2$ & 64 & - \\
6 & $Conv_{3}$ & $5\times5$ & 128 & $ReLU$ \\
7 & $MaxPooling_{3}$ & $2\times2$ & 128 & - \\
8 & $FC$ & - & 1024 & $ReLU$ \\
9 & $FC$ & - & 1024 & $ReLU$ \\
10 & $Output$ & $1\times2$ & - & $Softmax$ \\
\hline
\end{tabular}
\end{center}
\end{table}

\section{Searching for new M31 clusters}\label{sec:search}
\subsection{Data pre-processing and model training}\label{sec:preprocessing}

\begin{figure}
\center{
\includegraphics[scale=0.6]{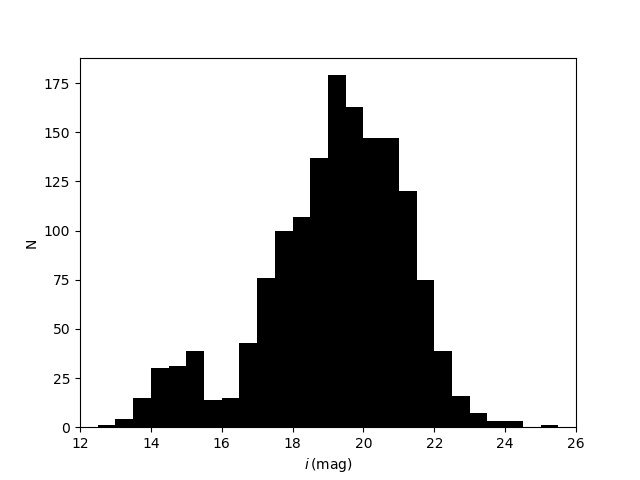}
}
\caption{$i$-band magnitude distribution of confirmed M31 clusters in our positive training sample.}
\label{fig:magnitude}
\end{figure}

Our candidate clusters were selected from all objects detected by the PAndAS, including those classified as point sources (with a  negative morphology flag), extended sources (with a positive morphology flag) and noise sources (with a morphology flag of zero) in the PAndAS source catalogue. This is because that there are confirmed M31 clusters from our positive sample in all the three categories. We selected object that have $13<i<23$\,mag. This range in magnitude encompasses the range of the confirmed M31 clusters (see Fig.~\ref{fig:magnitude}). This yielded 21,245,632 sources as our input sample.

We aim here to search for new M31 clusters from the combined $g$- and $i$-band image stamps of over 21 million sources from PAndAS by CNN. The input of machine-learning algorithms should be numerical. We therefore transferred the PAndAS $g$- and $i$- images into $56\times56\times2$ {\sc python} numpy float arrays. To make the training procedure fast and efficient, the input images were then normalised by constraining the pixel values into a small and symmetric range. Similar to \citet{2019MNRAS.483.4774L}, we adopted the standard z-score method, which is defined as
\begin{equation}
\begin{aligned}
Z=\frac{M-\mu}{\sigma},
\end{aligned}
\end{equation}
where $M$ represents the flux value of the individual pixel, $\mu$ is the mean flux value of the image stamp, and $\sigma$ is the standard deviation. The standard z-score is a commonly adopted method in machine learning and is highly efficient. The standard z-score preprocessing was applied to our training sample and to the input sample.
%In Fig.~\ref{fig:standardized} we show the comparisons of images before and after standardization. No visual difference is visible in the Figure. {\bf The purpose of inputted image standardization is to make the training procedure fast and efficient, by constraining the image pixel values in a small and symmetric range. But a correct standardization must protect the invariance of morphology of the image, so that no visual difference before and after the standardization is necessary and reasonable.}

%\begin{figure*}
%\center{
%\includegraphics[scale=0.5]{origin1.jpg}
%\includegraphics[scale=0.5]{normalized1.jpg}
%\includegraphics[scale=0.5]{origin2.jpg}
%\includegraphics[scale=0.5]{normalized2.jpg}
%}
%\caption{Left panels are original images of two sources randomly selected from positive training sample, while the right panels are corresponding normalized images. All images are in $i$-band.}
%\label{fig:standardized}
%\end{figure*}

In the beginning of the training, we labelled the positive and negative samples with the scheme called 'one hot coding' \citep{2015Packt Publishing}, which is the most widely used coding scheme to pre-process categorical attributes.We randomly divided the training sample (both positive and negative) into two parts, 80\%\ of the sources to train the model, and 20\%\  to test the training accuracy.

\begin{figure}
\center{
\includegraphics[scale=0.6]{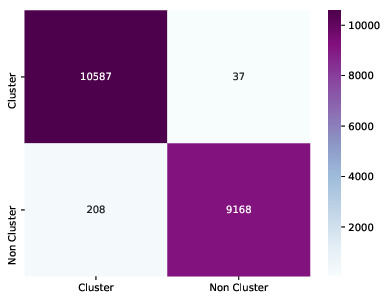}
}
\caption{Confusion matrix of the CNN for the binary classification. }
\label{fig:confusion}
\end{figure}

We applied our CNN to the training sample. To examine the accuracy of the training model, we adopted the F1 score \citep{1998Information Retrieval} and confusion matrix, which are excellent indicators of the accuracy of binary classification models. An F1 score value of 1 denotes the best accuracy of training, and 0 represents the worst. It is defined as
\begin{equation}
\begin{aligned}
F_{1}=2\cdot\frac{P\cdot R}{P+R},
\end{aligned}
\end{equation}
where $P$ presents the precision and $R$ the recall. The precision $P$ and recall $R$ are defined as
\begin{eqnarray}
P=\frac{\rm TP}{\rm TP+FP}, \\
R=\frac{\rm TP}{\rm TP+FN},
\end{eqnarray}
where TP means that positive classes are predicted as positive classes, while FP means that negative classes are predicted as positive classes, and FN denotes the false negative, which predicts the positive class as a negative class. We show in Fig.~\ref{fig:confusion} the confusion matrix of our classification results. From the confusion matrix we can obtain that our CNN classification model has a precision$=0.996$, a recall$=0.980$, an F1 score$=0.988$, and an accuracy$=0.988$.

\subsection{CNN prediction}

The trained CNN classification model was applied to all 21 million sources in our input sample. To achieve high confidence, we confirmed an CNN experimental candidate with its prediction probability corresponding to a positive sample higher than 0.95. Our model predicted 65,548 candidates in all observational fields. We made a quick scan of the images of these candidates. These candidates are all extended sources. Point sources are well excluded by the CNN model. However, the size of the sample of the predicted possible candidates was still too large, and some elliptical galaxies contaminated the sample. The contaminations were mainly because the negative sample did not contain enough galaxies. To reduce the resulting sample size and exclude the extragalactic contamination, we processed a second CNN classification using the same network and the same positive training sample, but a different negative training sample. We cross-matched the 65,548 candidates predicted from the first CNN model with the SIMBAD database, which yielded 4,363 sources classified as galaxies. The images of these galaxies were augmented in part with the same augmentation procedure as for the positive training sample (see Sec.~\ref{sec:augmentation}). This yielded a negative training sample of 38,592 galaxies. Together with the previous positive sample, they were adopted to train the second CNN classification model. Again, 80\%\  of the positive and negative training sample were adopted for training and 20\%\  to test the accuracy. We obtained an F1 score$=0.980$ and an accuracy$=0.981$ for the second CNN network.

The second network was then applied to the 65,548 candidates predicted by the first CNN model. A prediction probability threshold of 0.98 was adopted to obtain the final CNN output, which yielded 5,092 M31 star cluster candidates. We cross-matched these candidates with the SIMBAD database and obtained 523 objects. Three hundred and eighty-four of the 523 objects are classified as possible clusters, which are marked as C?* (possible (open) star cluster), Gl? (possible globular cluster), Cl* (cluster of stars), or GlC (globular cluster). The remaining 139 objects are galaxies, a galaxy in a cluster of galaxies, an X-ray source, an infra-red source, a radio source, and so on. We visually inspected the PAndAS images of these 523 SIMBAD objects. Except for the confirmed clusters, none of the objects can be classified as bona fide clusters. They were thus excluded from our sample. We also cross-matched our candidates with catalogues from the literature (e.g. \citealt{2008ApJS..177..174N}) and found no counterparts.

\subsection{Visual inspection}

\begin{figure}
\center{
\includegraphics[scale=0.6]{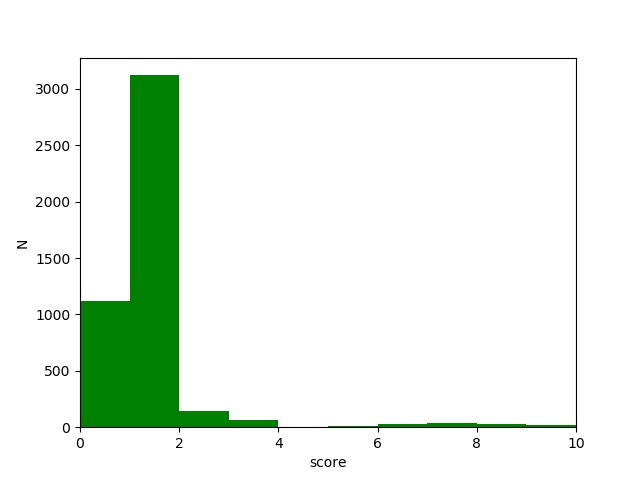}
}
\caption{Histogram of the summed scores of the CNN-predicted candidates from all inspectors.}
\label{fig:scores}
\end{figure}

Finally, the 4,569 remaining M31 cluster candidates predicted by the second CNN were independently examined by five experienced human inspectors through visual inspection. Each inspector handed in the rating scores, which were defined as follows: 2 points for $sure$ star clusters, 1 points for $maybe$ star clusters, and 0 points for $non$-star clusters. In Fig.~\ref{fig:scores} we show the distribution of the summed scores of all CNN candidates from the inspectors. Scores of 0, 1, 2, 3, 4, 5, 6, 7, 8, 9, and 10 were allotted to  1117, 3121, 148, 64, 2, 7, 24, 39, 27, 14, and 6 sources, respectively. Adopting the threshold 5 for the summed scores, we finally obtained 117 objects as high-probability M31 star cluster candidates.

\section{Results and discussion}\label{sec:result}

\begin{figure*}
\center{
\includegraphics[scale=0.4]{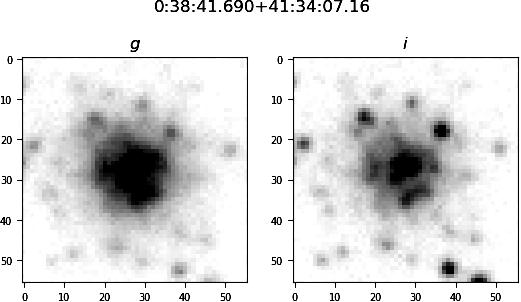}
\includegraphics[scale=0.4]{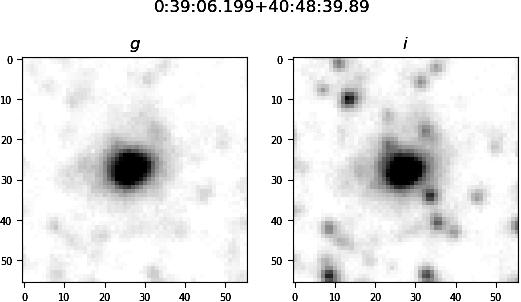}\\
\vbox{}
\includegraphics[scale=0.4]{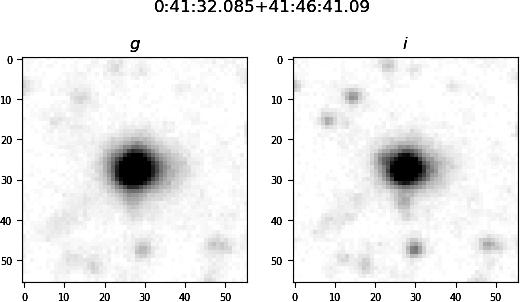}
\includegraphics[scale=0.4]{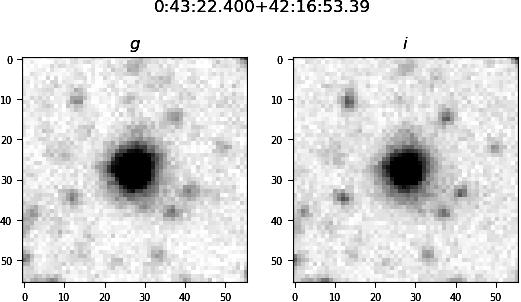}\\
\vbox{}
\includegraphics[scale=0.4]{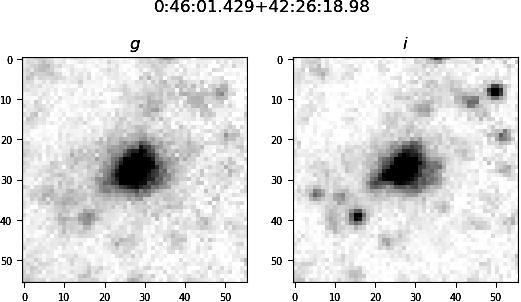}
\includegraphics[scale=0.4]{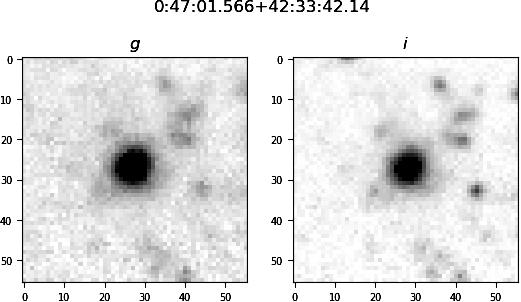}
}
\caption{PAndAS $g$- and $i$-band images of the six new M31 cluster candidates that have the highest summed score from the inspectors. The coordinates of the objects are labelled.}
\label{fig:candidate}
\end{figure*}

The 117 new M31 cluster candidates are listed in Table~\ref{tab:coordinate}. Each row of Table~\ref{tab:coordinate} contains the coordinates (RA and Dec), PAndAS $g$- and $i$-band magnitudes, projected distances to the centre of M31 $R_{\rm proj}$, ellipticity $e$, position angle (P.A.), and the summed score from the five inspectors of one identified M31 cluster candidate. The PAndAS $g$- and $i$-band magnitudes were adopted from \citet{2018ApJ...868...55M}. The values of $e$ and P.A. of the individual clusters were calculated from the PAndAS $i$-band image with the {\sc stsdas} {\sc ellipse} task \citep{2012ascl.soft06003S}. The catalogue is available in electronic forms at the CDS. In Fig.~\ref{fig:candidate} we show the PAndAS $g$- and $i$-band negative images with enhanced contrast of the six new M31 cluster candidates that have the highest summed score (10) from all the inspectors. Their morphological appearances are similar to those of the confirmed M31 clusters.

\begin{figure}
\center{
\includegraphics[scale=0.6]{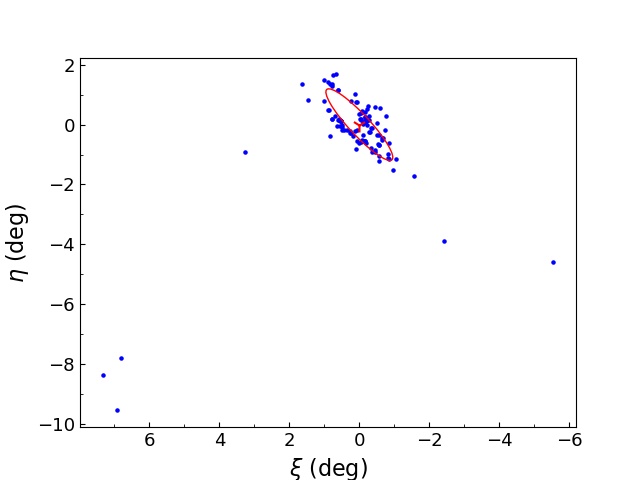}
}
\caption{Spacial distribution of the newly identified 117 M31 cluster candidates. The coordinate (0,0) represents the centre of M31 (with a red `Y'; RA=$00^{\rm h}42^{\rm m}44^{\rm s}.30$, Dec=$+41\degr16^{\prime}09^{\prime\prime}.0$).  The red ellipse represents the optical disk of M31, for which we adopt a major axis length of 1.59\degr, a disk inclination angle of 78\degr, and a major axis position angle of 38\degr.}
\label{fig:distribution}
\end{figure}

The spatial distribution of the new cluster candidates is shown in Fig.~\ref{fig:distribution}. Eight cluster candidates are located in the halo of M31, with $R_{\rm proj}>$25\,kpc. The farthest candidate is about 158\,kpc away from the centre of M31. Most of the new cluster candidates located in the disk of M31 are located southeast of M31. This is because the disk clusters we identified are mainly young clusters. The disk northwest of M31 was observed by the PHAT project \citep{2012ApJS..200...18D}; the young clusters were previously identified \citep{2012ApJ...752...95J,2015ApJ...802..127J}.

\begin{table*}
\caption{Catalogue of the 117 new M31 star cluster candidates.}
\label{tab:coordinate}
\begin{center}
\begin{tabular}{ccccccccc}
  \hline
  \hline
Name & RA (J2000) & Dec (J2000) & $g$ (mag) & $i$ (mag) & $R_{\rm proj}$ (kpc) & $e$ & P.A. (degree) & Score\\
\hline
Candidate1 & 0:15:05.956 & +36:27:38.25 & 21.856 & 20.308 & 96.7 & 0.08 & 15.5 & 6 \\
Candidate2 & 0:30:32.016 & +37:20:41.34 & 21.133 & 19.865 & 61.5 & 0.18 & 11.8 & 6 \\
Candidate3 & 0:34:42.035 & +39:31:24.42 & 20.771 & 19.996 & 31.2 & 0.07 & $-$21.13 & 8 \\
Candidate4 & 0:37:16.566 & +40:06:36.16 & 20.949 & 19.984 & 20.9 & 0.16 & 18.39 & 6 \\
Candidate5 & 0:37:48.369 & +39:44:29.92 & 21.991 & 21.156 & 24.1 & 0.32 & $-$27.48 & 7 \\
...... & ...... & ...... & ...... & ...... & ...... & ...... & ...... & ...... \\
\hline
\end{tabular}
\end{center}
\footnotesize{$^a$ The full table is available at the CDS and also via the website \url{http://paperdata.china-vo.org/Wang.SC/2021/table2.fits}.}
\end{table*}

\subsection{Morphology}

\begin{figure*}
\center{
\includegraphics[scale=0.4]{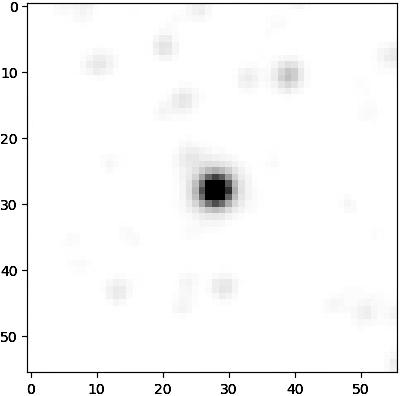}
\includegraphics[scale=0.4]{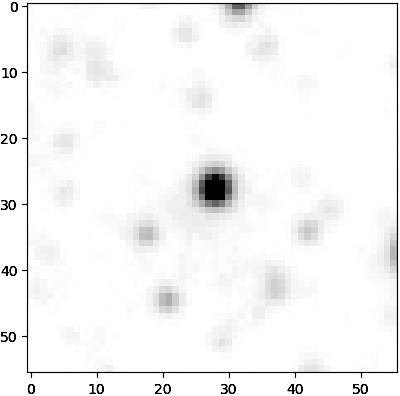}
\includegraphics[scale=0.4]{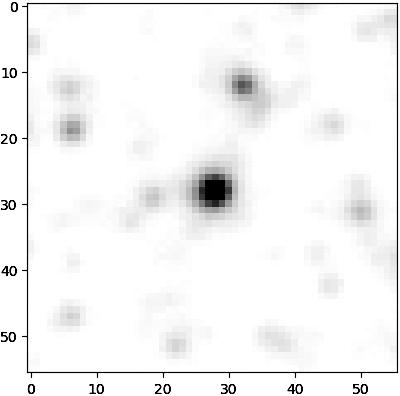}
\includegraphics[scale=0.4]{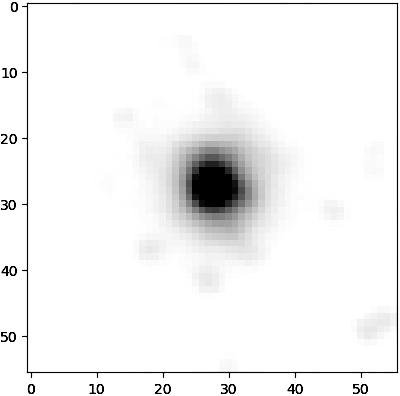}
\includegraphics[scale=0.4]{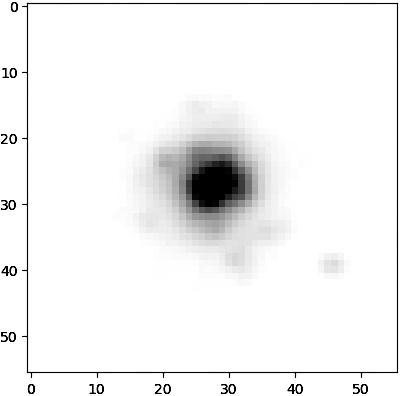}
\includegraphics[scale=0.4]{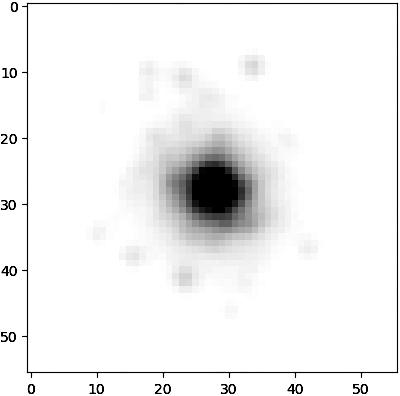}
%\begin{overpic}[scale=0.38]{johnson8.txt.jpg}\put(51,23){\includegraphics[scale=0.28]{johnson8.jpg}}\end{overpic}
%\begin{overpic}[scale=0.38]{johnson17.txt.jpg}\put(51,23){\includegraphics[scale=0.28]{johnson17.jpg}}\end{overpic}
%\begin{overpic}[scale=0.38]{johnson18.txt.jpg}\put(51,23){\includegraphics[scale=0.28]{johnson18.jpg}}\end{overpic}
%\begin{overpic}[scale=0.38]{PAndAS-27.txt.jpg}\put(51,23){\includegraphics[scale=0.28]{PAndAS-27.jpg}}\end{overpic}
%\begin{overpic}[scale=0.38]{PAndAS-04.txt.jpg}\put(51,23){\includegraphics[scale=0.28]{PAndAS-04.jpg}}\end{overpic}
%\begin{overpic}[scale=0.38]{PAndAS-37.txt.jpg}\put(51,23){\includegraphics[scale=0.28]{PAndAS-37.jpg}}\end{overpic}
}
\caption{PAndAS $i$-band images of six example confirmed clusters. The three clusters in the upper panels are young disk clusters selected from \citet{2012ApJ...752...95J}, and those in the bottom panels are globular clusters from \citet{2014MNRAS.442.2165H}.}
\label{fig:sample-morphology}
\end{figure*}

\begin{figure*}
\center{
\includegraphics[scale=0.3]{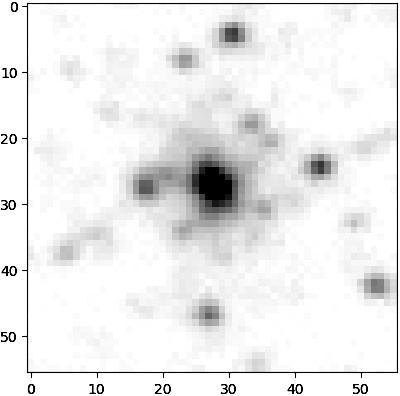}
\includegraphics[scale=0.3]{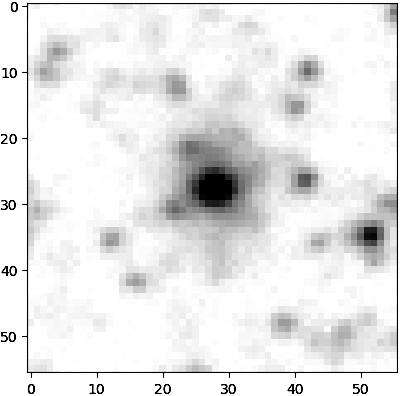}
\includegraphics[scale=0.3]{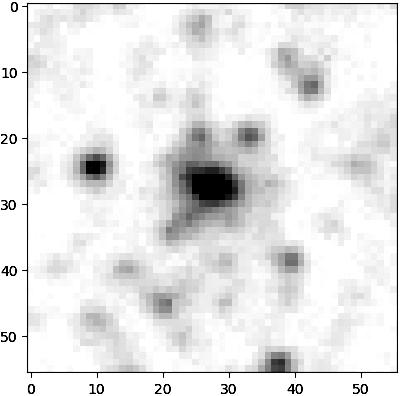}
\includegraphics[scale=0.3]{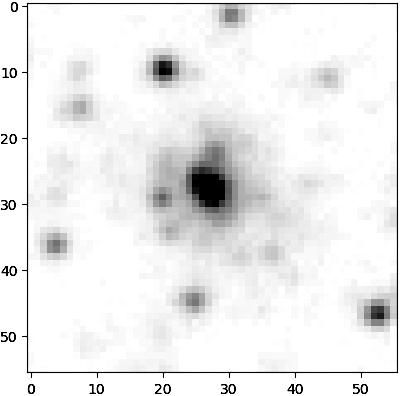}
\includegraphics[scale=0.3]{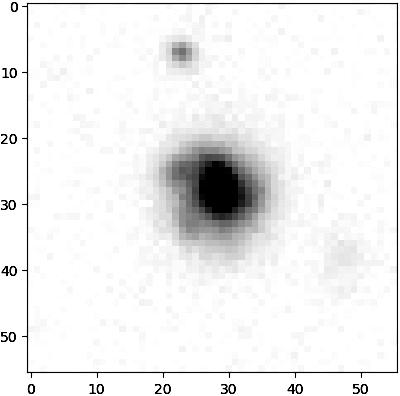}
\includegraphics[scale=0.3]{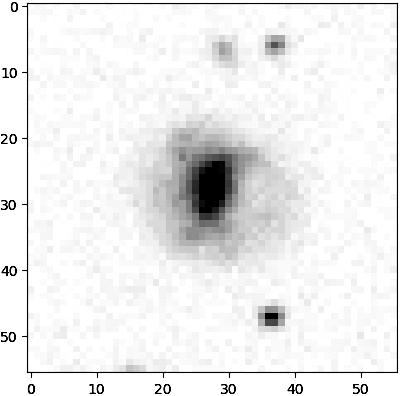}
\includegraphics[scale=0.3]{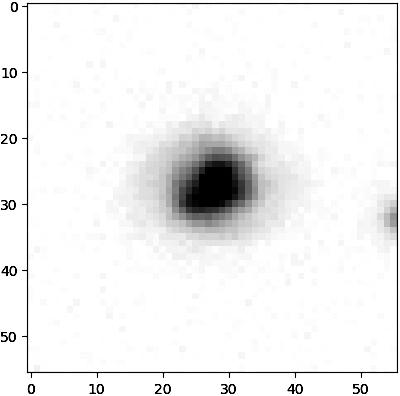}
\includegraphics[scale=0.3]{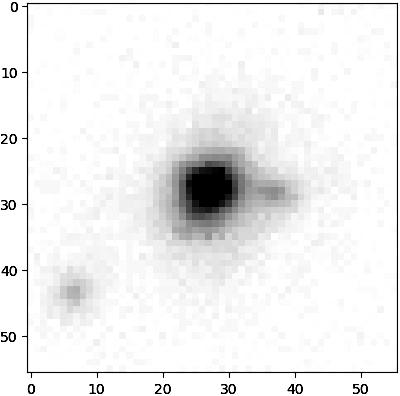}
\includegraphics[scale=0.3]{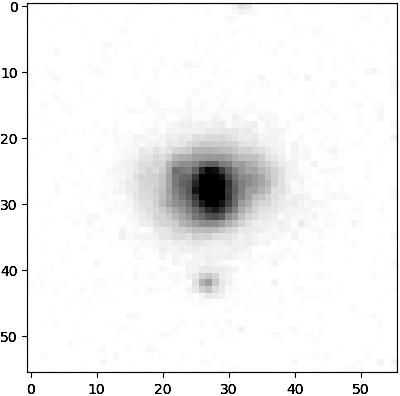}
\includegraphics[scale=0.3]{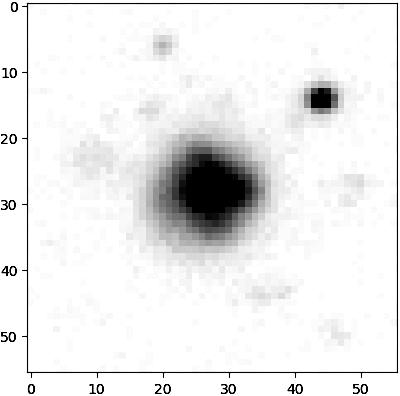}
\includegraphics[scale=0.3]{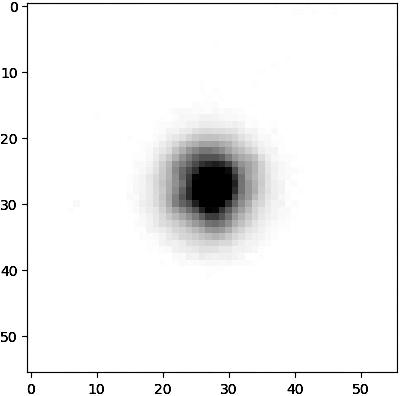}
\includegraphics[scale=0.3]{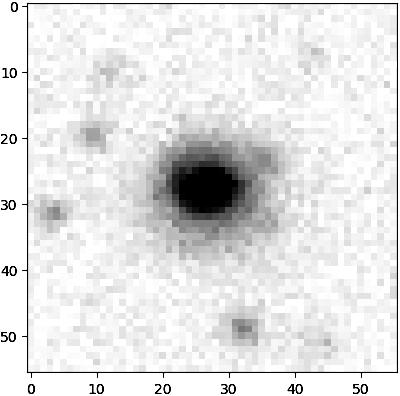}
}
\caption{Similar as in Fig.~\ref{fig:sample-morphology}, but for the 12 example cluster candidates that we newly identified. The four candidates in the first row are young disk cluster candidates, and the other eight objects in the second and third rows are the eight halo candidates with $R_{\rm proj}>$25\,kpc.}
\label{fig:candidates-morpholoy}
\end{figure*}

Here we confirm the morphological similarity between our newly identified M31 cluster candidates and the confirmed clusters.
In Fig.~\ref{fig:sample-morphology} we show the morphology of six example M31 clusters from our positive training sample. The three example clusters plotted in the upper panel of Fig.~\ref{fig:sample-morphology} are young disk clusters selected from \citet{2012ApJ...752...95J}, and the other three are old halo clusters from \citet{2014MNRAS.442.2165H}. Overall, young disk clusters tend to be more symmetric and compact. In Fig.~\ref{fig:candidates-morpholoy} we show the morphology of 12 example M31 cluster candidates that we newly identified. Most of our candidates are located southeast of the M31 disk, with projected radii $R_{\rm proj}<$20\,kpc. We suggest that these candidates are young disk clusters. The first row of Fig.~\ref{fig:candidates-morpholoy} shows four example clusters that are located in the disk. Their morphology is analogous to that of the young disk clusters from \citet{2012ApJ...752...95J}. We also identified eight new clusters in the halo of M31 (with $R_{\rm proj}>$25\,kpc). They are also shown in Fig.~\ref{fig:candidates-morpholoy} and exhibit similar features as the globular clusters from \citet{2014MNRAS.442.2165H}.

\subsection{Radial distribution}
\begin{figure}
\center{
\includegraphics[scale=0.6]{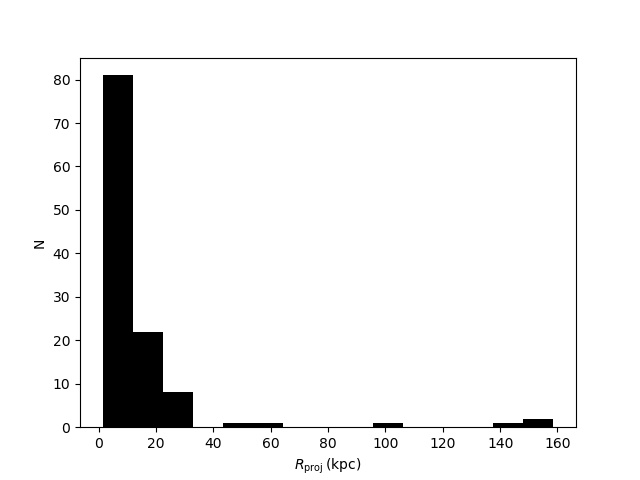}
}
\caption{Histogram of the projected distance to the M31 centre of all the newly identified M31 cluster candidates.}
\label{fig:ra-dist}
\end{figure}

Fig.~\ref{fig:ra-dist} show the radial distribution of our newly identified 117 M31 cluster candidates. Most of the clusters are located in the disk of M31 and are concentrated in the area with galactocentric radii $R_{\rm proj}<$10\,kpc. They are mainly young clusters with a morphology analogous to those from \citet{2012ApJ...752...95J}. Eight clusters are located in the halo of M31. Three of them have projected radii larger than 100\,kpc. They are all old clusters and have an extended radial profile, according to their morphology.

\subsection{Colour-magnitude diagram}
\begin{figure}
\center{
\includegraphics[scale=0.6]{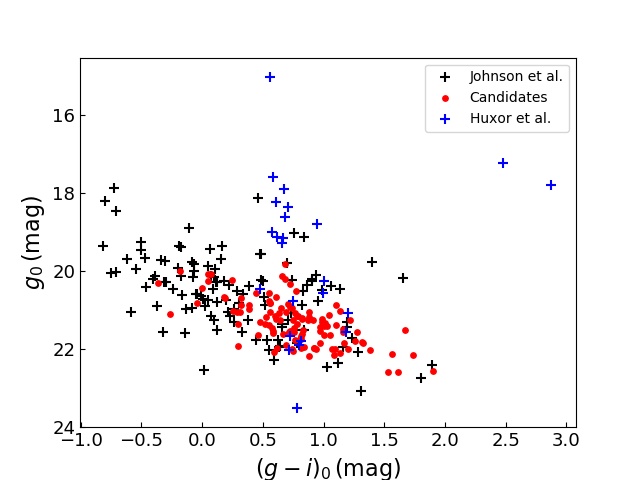}
}
\caption{Colour-magnitude diagram of the newly identified M31 cluster candidates (red circles) and the confirmed clusters from \citet[][black pluses]{2012ApJ...752...95J} and \citet[][blue pluses]{2014MNRAS.442.2165H}, respectively.}
\label{fig:color-mag}
\end{figure}

In Fig.~\ref{fig:color-mag} we plot the colour-magnitude diagram of all our newly identified M31 cluster candidates. The young disk clusters from \citet{2012ApJ...752...95J} and halo globular clusters from \citet{2014MNRAS.442.2165H} are overplotted for comparison. We adopted the PAndAS $g$ and $i$-band magnitudes. Similar as in \citet{2021A&A...645A.115W}, the reddening values of the individual objects were either adopted from the literature or the median reddening values of all the known clusters within 2\,kpc. Significant discrepancies are visible between the young disk clusters from \citet{2012ApJ...752...95J} and the halo globular clusters from \citet{2014MNRAS.442.2165H}. Young clusters are systematically bluer and fainter. Our newly identified M31 clusters candidates have similar colours and magnitudes as the Johnson et al. young clusters, which suggests that most of them are young disk clusters.

%\begin{figure*}
%\center{
%\includegraphics[scale=0.35]{3350.jpg}
%\includegraphics[scale=0.35]{2917.jpg}
%\includegraphics[scale=0.35]{3308.jpg}
%\includegraphics[scale=0.35]{3503.jpg}
%\includegraphics[scale=0.35]{3855.jpg}
%\includegraphics[scale=0.35]{3841.jpg}
%}
%\caption{Isobrightness lines (in red) in $i$-band corresponding to those six maximum-score candidates shown in Fig.~\ref{fig:candidate}. }
%\label{fig:isobrightness}
%\end{figure*}
%
%\begin{figure*}
%\center{
%\includegraphics[scale=0.5]{3350.eps}
%\includegraphics[scale=0.5]{2917.eps}
%\includegraphics[scale=0.5]{3308.eps}
%\includegraphics[scale=0.5]{3503.eps}
%\includegraphics[scale=0.5]{3855.eps}
%\includegraphics[scale=0.5]{3841.eps}
%}
%\caption{Same as Fig.~\ref{fig:isobrightness}, but the surface brightness profiles. Black circles represent intensities (in the from of pixel value, with cyan error bars).}
%\label{fig:profile}
%\end{figure*}

\section{Summary}\label{sec:summary}
We have constructed double-channel CNN classification models to search for new M31 star clusters. Confirmed M31 clusters and non-cluster objects from the literature were selected as our training sample. The trained CNN networks were applied to the $g$- and $i$-band images of over 21 million sources from PAndAS, which have predicted 4,569 machine-learning candidates. After human inspection, we identified 117 new high-probability M31 cluster candidates. Most of our candidates are young clusters and are located in the disk of M31. Eight candidates are located in the halo of M31, with $R_{\rm proj}>$25\,kpc. The farthest candidate has a projected distance of $R_{\rm proj}=$158\,kpc. We also compared the morphology, colours, and magnitudes of our newly identified clusters with those of the young disk clusters from \citet{2012ApJ...752...95J} and halo globular clusters from \citet{2014MNRAS.442.2165H}. Obvious systematic discrepancies are visible between the young and old clusters. The features of our new candidates that are located in the disk are similar to those of the young clusters from \citet{2012ApJ...752...95J}, while the eight candidates located in the halo of M31 exhibit similar features as the globular clusters from \citet{2014MNRAS.442.2165H}.

Contaminants such as M31 HII regions, background galaxies, and foreground stars may exist in our catalogue. Spectroscopic observations of these candidates for a final identification in the future are required.

\section*{Acknowledgement}
 This work is partially supported by National Key R\&D Program of China No. 2019YFA0405501 and 2019YFA0405503, National Natural Science Foundation of China (NSFC) No. 11803029, 11873053 and 11773074, and Yunnan University grant No.C176220100007. This work has made use of data products from the Pan-Andromeda Archaeological Survey (PAndAS). PandAS is a Canada-France-Hawaii Telescope (CFHT) large program that was allocated 226 hours of observing time on MegaCam. This work also has utilized the data (catalogues) from the Guoshoujing Telescope (the Large Sky Area Multi-Object Fibre Spectroscopic Telescope, LAMOST), Gaia Data Release 2 and websites by Nelson Caldwell.

\end{document}